\documentclass[twocolumn,prd,showkeys,showpacs]{revtex4}
\usepackage{latexsym} 
\usepackage{amssymb} 
\usepackage{graphicx}
\newcommand{\Tr}{\textrm{Tr}\,}
\newcommand{\avr}[1]{\left\langle{#1}\right\rangle}
\newcommand{\p}{{\vec p}}
\newcommand{\q}{{\vec q}}
\newcommand{\x}{{\vec x}}
\newcommand{\y}{{\vec y}}
\newcommand{\twomatrix}[1]{\left(\begin{array}{cc}#1\end{array}\right)}
\newcommand{\twovector}[1]{\left(\begin{array}{c}#1\end{array}\right)}

\renewcommand{\Re}{\textrm{Re}\,}
\renewcommand{\Im}{\textrm{Im}\,}
\newcommand{\intp}{\int_{\vec p}}

\newcommand{\anticomm}[2]{\left\{#1,#2\right\}}
\newcommand{\comm}[2]{\left[#1,#2\right]}
\newcommand{\evalat}[2]{\left.#1\right|_{#2}}
\newcommand{\pabs}{|\vec p|}
\newcommand{\phat}{{\hat p}}

\newcommand{\phiu}{\phi^{{\rm u},s}(x,\vec p)}
\newcommand{\phiv}{\phi^{{\rm v},s}(x,\vec p)}
\newcommand{\phivm}{\phi^{{\rm v},s}(x,-\vec p)}
\newcommand{\omegap}{\omega_{\vec p\,}}
\newcommand{\nn}{\nonumber\\}
\begin{document}
\title{Low-cost fermions in classical field simulations}
\author{ Sz. Bors\'anyi}
\email{s.borsanyi@sussex.ac.uk}
\author{M. Hindmarsh}
\email{m.b.hindmarsh@sussex.ac.uk}
\affiliation{
Department of Physics and Astronomy,  University of Sussex,
Brighton, East Sussex BN1 9QH, United Kingdom.
}
\date{\today}
\begin{abstract}
We discuss the possible extension of the bosonic classical field theory
simulations to include fermions. This problem has been addressed in terms of
the inhomogeneous mean field approximation by Aarts and Smit.  By performing a
stochastic integration of an equivalent set of equations we can extend the
original 1+1 dimensional calculations so that they become feasible in higher
dimensions. We test the scheme in 2 + 1 dimensions  and discuss some classical
applications with fermions for the first time, such as the decay of oscillons.
\end{abstract}
\pacs{03.65.Pm, 11.10.-z, 11.15.Kc}
\keywords{Classical field theory; Lattice fermions; Oscillons; Hartree approximation}
\maketitle
\section{\label{sec:intro}Introduction}

Since the advent of modern computational facilities classical field
theory is one of the most popular approaches to nonequilibrium field theory.
The classical approximation to a quantum filed theory is well justified in
several cosmological applications ranging form reheating of the
postinflationary Universe \cite{ReheatingUS,ReheatingUSSR} followed by an
evolution through various phase transitions \cite{PhaseTransitions} to the
nonlinear evalution of the hypothetical cosmic strings
\cite{EnglishStrings,VilenkinShellard,RussianStrings}.
Classical methods have also received an increasing amount of attention
from the heavy ion community.  The initial evolution of the highly
excited gluon plasma in little-bang experiments turns out to be well modelled
by classical Yang-Mills equations \cite{InitialKrasnitzVenugopalan}. 

The preheating of the inflationary Universe was one of the pioneering
applications of the nonlinear classical wave equations
\cite{ClassicalReheating}. In the mostly studied chaotic and hybrid inflation
scenarios the nonlinear dynamics is driven by an instability, which is
parametric or tachyonic, respectively.  Instabilities lead to nonperturbatively
large occupation numbers, which is a prerequisite for the classical
approximation, but it also requires a nonperturbative treatment, which is the
actual strength of the classical equations.  The classical simulations of
preheating can make estimates on non-gaussian density perturbations
\cite{ChambersRajantie} and gravitational wave production
\cite{GravityWavesPreheating,GravityWaveProd,GravityWavesHybrid}.

Nonperturbative methods are especially useful when dealing with nonequilibrium
phase transitions in the Early Universe \cite{PhaseTr}. A typical example
where the fields are required to be out of equilibrium is baryogenesis.  A
second strength of the classical approximation is that equilibrium is not a
prerequisite.  Solving the real time Yang-Mills equations with
far-from-equilibrium initial conditions one could gain access to the evolution
of the Chern-Simons number
\cite{EwBaryogenesis,EwBaryogenesisMoore,EwBaryogenesisUK}.  For this to
accomplish the third strength of the classial equations has been exploited: 
its preservation of gauge invariance under time-independent transformations.

The fourth strength of the classical approach is its simplicity and cheap
implementation even at large scales. This feature makes it an excellent tool
for studying topological defects, especially the hypothetical network of cosmic
strings. To
address formation and evolution of defect networks very different length scales
have to be properly incorporated into one numerical computation. This situation
is getting worse in an expanding universe, but in the classical setting these
calculations are still affordable \cite{ExpandingStrings}. In principle, one
could take the zero-width limit and solve the Nambu-Goto equations
\cite{NambuGoto}. For fundamental strings this is a natural procedure, but for
strings which are topological defects, microscopic physics plays a significant
role in the decay mechanism of strings \cite{ClassicalStrings1}.
Explicit calculations have been made in the context of gauge strings in the
Abelian Higgs model
\cite{ClassicalStrings1,ClassicalStrings2,ExpandingStrings}, global
\cite{GlobalStrings} and semilocal strings
\cite{SemilocalStrings,SussexStringsinCMB2} as well as domain walls
\cite{ClassicalWalls1,ClassicalWalls2}. The relevance of these calculations
have been recently highlighted by the discovery of possible traces of cosmic
strings in the cosmic microwave background \cite{SussexStringsinCMB1,
SussexStringsinCMB2}.

Besides of the cosmological interest classical field theory simulations are
also used on the subatomic scale: the early evolution of the gluon plasma formed in 
heavy ion collisions can be described by 
by classical Yang-Mills equations \cite{InitialKrasnitzVenugopalan}.  This
facilitates a nonperturbative description of the glasma, i.e.  the intermediate
state after the melting of the color glass condensate prior to thermalisation
to quark gluon plasma \cite{GlasmaLappi,GlasmaVenugopalan}.  The produced
nonabelian plasma is highly anisotropic, and as such, it is subject to
instabilities \cite{PlasmaInst}. Classical methods have proved very
useful for giving a quantitative account on the isotropisation driven by these
Weibel instabilities \cite{GlasmaRomatschke,DarmstadtClassicalYM}.
Alternatively, one can replace the hard sector of the field theory by classical
particles represented by a set of Vlasov equations on the background of soft
classical fields \cite{VlasovAM,VlasovBodeker}.

Finally we point out that the ergodicity of the classical field trajectories
makes the classical simulations an essential and robust method for studying
thermal classical lattice systems in real time. In statistical field theory one
averages over an ensemble of initial field configurations and observes e.g. the
real time dynamics of symmetry breaking \cite{SzepThesis},
with possible formation of quasistable localised excitations, dubbed oscillons
\cite{GleiserOscillon}.  The presence of long-lived
oscillons induces resonant nucleation, and they become a driving force of
first order phase transitions \cite{GleiserBubble}.

The classical approximation has severe limitations, however. The continuum
equilibrium theory is plagued by Rayleigh-Jeans divergences, and a
renormalisation with local counterterms is not possible in general
\cite{ClassicalRenorm}. Moreover, the counterterms are temperature dependent,
which makes a consistent out-of-equilibrium renormalisation impossible.
This means that classical theories need an intrinsic cut-off scale,
which, in practice, sets the spacing of the lattice discretisation.
From whatever initial ensemble of classical fields the
straightforward integration of the Euler-Lagrange equations of the theory
brings the systems towards an equilibrium defined by the classical Hamiltonian.
This equilibrium differs from a true quantum thermal state, but the difference
is negligible for soft modes and only affects hard excitations. In terms of
particle numbers a system is considered in the classical domain if the
occupancy is sufficiently high. The infrared physics, which is mostly sensitive
to nonperturbative phenomena, is usually not vulnerable to quantum effects, but
on the ulraviolet end of the spectrum one has to balance between discretisation
errors and miscalculated hard degrees of freedom. Even if we start from an
infrared dominated initial condition, hard modes are becoming increasingly
dominant on the course of thermalisation and the classical system automatically
leaves its domain of validity.

There is an other first principles approach to nonequilibrium field theory,
which shares none of the aforementioned shortcomings. It has been numerically
demonstrated that even a low order truncation of the two-particle irreducible
(2PI) effective action yields equations of motion, capable of describing
irreversible quantum dynamics, including thermalisation \cite{Jurgen}.  This
powerful resummation technique can be directly applied to relevant problems in
cosmology \cite{Berges:2002cz,Berges:2008wm} or in hot abelian gauge theories
\cite{AartsQED,BorsanyiQED}, as well as in the many-body theory of ultracold
condensates \cite{Ultracold}. Yet, for non-abelian gauge fields the more
complete 3PI resummation becomes necessary \cite{LPM,ArnoldKinetic}, and for a
setting with topological defects an inhomogeneous treatment is inevitable
\cite{Rajantie:2006gy}.  Being both extensions expensive, we will have to fall
back in these cases to the classical approximation and use 2PI to benchmark it
where their domains of validity overlap. These precision tests in the $O(N)$
scalar model had the reassuring result that particle numbers as small as $\sim
10$ already put the system into the classical domain
\cite{ABclassical,AmsterdamTachyonic}.

There is, however, another important deficiency of the classical approximation.
All the applications listed in the previous paragraphs were entirely limited
to bosonic fields. Classical simulations of both baryogenesis and heavy
ion collisions could benefit from a direct modelling of quarks, if that
were feasable.

As dimensional reduction suggests, fermionic fields are
purely quantum degrees of freedom, just like the non-static components
of a bosonic field theory. 
The classical field theory does have bosonic fluctuations, and in the absence
of quantum degrees of freedom, bosonic particle production is automatically
modelled as the excitation of the fluctuating background. The analagous
production of fermions, however, is not mapped onto any existing degree of
freedom. 

The inclusion of fermions is rather trivial in the 2PI framework, where bosonic
quantum fluctiatons interact with fermionic quantum fluctuations, and an
explicit calculation has been presented to show the real-time simultaneous
onset of Fermi-Dirac and Bose-Einstein distributions \cite{Fermion2PI}. However,
when dealing with non-abelian gauge fields, or strong inhomogeneities, we will
need to resort to some extension of the classical theory. This extension is the
actual topic of this paper.

In this paper we build on the ideas of Aarts and Smit
\cite{AartsSmitNPB,AartsSmitPRD} and by ``integrating the fermion determinant''
we solve an effective theory for the classical scalar background.
We go beyond the recent applications in
Refs.~\cite{Gibbons:2006ge,TranbergSaffin} by including the back reaction
in our calculation. Our efficient solution technique enables us to go
beyond 1+1 dimensions in the simulations.

In Section \ref{sec:hartree} we review the standard description of the
fermionic fluctuations. Then in Section \ref{sec:stochastic} we introduce
a stochastic approach, which provides us a more efficient algorithm than
the so far known mode function expansion. In Section \ref{sec:therm}
we investigate the capabilities of this semiclassical approximation for
describing irreversible phenomena, such as damping and thermalisation.
We continue with a bit more exotic application involving oscillons
in Section \ref{sec:oscillon} and discuss the possible future applications
of this semiclassical scheme Section \ref{sec:discussion}.
The spinor representations that we actually used in our numerics we give
in Appendix~\ref{sec:spinors}. In a naively discertised lattice field theory
the number of fermion flavours is doubled in each space-time direction. We
discuss the possible elimination of the extra flavours in
Appendix~\ref{sec:doublers}.


\section{\label{sec:hartree}Integrating the fermion degree of freedom}

\subsection{\label{sec:model}A scalar model with fermions}
Let us pick a simple scalar model coupled to a fermion flavour through
Yukawa interaction:
\begin{eqnarray}
&&{\cal L}=\frac12\partial\Phi^*\partial\Phi-V(\Phi)\nonumber\\
&&\qquad+\sum_k \left[
i\bar\Psi_k \gamma^\mu\partial_\mu\Psi_k
-\bar\Psi_k(M P_L+M^* P_R)\Psi_k
\right]\quad
\end{eqnarray}
Here the $M$ complex fermion mass is a function of the background:
\begin{equation}
M(x)=m-g\Phi(x)\,.
\end{equation}
The projectors are defined as $P_L=\frac12(1-\gamma^5)$ and 
$P_R=\frac12(1+\gamma^5)$. The index $k$ runs over $N_f$
identical fermion flavours.
We will not use any of the special features of the bosonic sector,
and our discussion below will also apply to classical lattice gauge
theories with a covariant coupling to fermions.

Before going into details we summarise our strategy by defining a
bosonic effective action $\Gamma[\Phi]$ as
\begin{equation}
e^{i\Gamma[\Phi]}=\int \prod\limits_k D\Psi^+_k D\Psi_k e^{i\int{\cal L}(\Phi,\Psi^+,\Psi)}\,.
\label{eq:effpot}
\end{equation}
Our goal is to solve the semiclassical equation of motion
$\delta\Gamma[\Phi]/\delta\Phi=0$ without further approximation.
This path integral has to be understood on a real time contour with
a forward and backward time branch. To contour ends at (zero) initial
time where it connects to the initial density operator. We will
use a the perturbative vacuum or many-particle state as an initial
condition.

The Dirac equation written for the spinor operators is linear
\begin{eqnarray}
(i\gamma^{\mu}\partial_{\mu}-m+g\Re\Phi(x)-ig\Im\Phi(x)\gamma^5)\Psi(x)=0,
\label{eq:Dirac}&&\,\\
i\partial_{\mu}\bar\Psi(x)\gamma^{\mu}+\bar\Psi(x)(m-g\Re\Phi(x)+ig\Im\Phi(x)\gamma^5)=0,&&\,
\end{eqnarray}
which manifests on the level of diagrams in the simple rule that fermion
propagator lines never cross. The interaction is mediated
by the bosonic field, which is modelled by a fluctuating background.

Instead of using anticommuting operators we rewrite the Dirac equation
so that it applies to the symmetrised two-point function:
\begin{eqnarray}
&&(i\gamma^{\mu}\partial_{x,\mu}-m\nn
&&\quad+g\Re\Phi(x)-ig\Im\Phi(x)\gamma^5)D(x,y)=0,\label{eq:Dx}\\
&&i\partial_{y,\mu}D(x,y)\gamma^{\mu}+\nn
&&\quad D(x,y)(m-g\Re\Phi(y)+ig\Im\Phi(y)\gamma^5)=0\label{eq:Dy}\,,
\end{eqnarray}
where $D(x,y)$ is defined as
\begin{eqnarray}
D(x,y)_{ij}&=&
\frac12\left(D^>_{ij}(x,y)-D^<_{ij}(x,y)\right)\nn
&=&\frac12\avr{\comm{\Psi_i(x)}{\bar\Psi_j(y)}}\,,
\label{eq:Ddef}\\
D^<(x,y)_{ij}&=&\left\langle \bar\Psi_j(y) \Psi_i(x)\right\rangle\,,\\
D^>(x,y)_{ij}&=&\left\langle \Psi_i(x)\bar\Psi_j(y)\right\rangle\,.
\end{eqnarray}
where $i,j$ represent the Dirac as well as flavour indices. The propagator
$D$ is identical to the $F$-type two-point function in the literature
of nonequilibrium Green's functions as well as in Ref.~\cite{Fermion2PI}.
One can work out an equation for the spectral function as well, which will
take an identical form.

The bosonic background obeys a simple wave equation,
\begin{equation}
\partial^2_x\Phi(x)+V'(\Phi(x))+N_f J(x)=0\,
\label{eq:Phi}
\end{equation}
where the fermionic back reaction is carried by the current $J$, which is a
combination of the scalar and pseudoscalar currents:
\begin{eqnarray}
J(x)&=& J^{\rm S}(x) + J^{\rm PS}(x)=2g\Tr D(x,x) P_R\,,\label{eq:J}\\
J^{\rm S}(x)&=&-g\avr{\bar\Psi(x)\Psi(x)}=g\Tr D(x,x)\,,
\label{eq:JS}
\\
J^{\rm PS}(x)&=&-g\avr{\bar\Psi(x)\gamma^5\Psi(x)}=g\Tr D(x,x)\gamma^5\,.
\label{eq:JPS}
\end{eqnarray}
The scalar current is always real the pseudoscalar current is always imaginary.

In a theory with a Dirac mass $m$ the vacuum propagator takes the following
form
\begin{equation}
\left.D(x^0,\vec x,y^0,\vec y)\right|_{x^0=y^0}
=\intp e^{-i p_j(x^j-y^j)}
\frac{m+p_i\gamma^i}{2\omega_\p}\,,
\label{eq:Dic}
\end{equation}
with $\omegap^2=m^2+\pabs^2$. The latin indices refer to space only. 
In this equation we introduced the notation $\intp$ for the three dimensional
momentum integral $\int{d^3p}/{(2\pi)^3}$. 

In many cases when one inquires about the fermion production the vacuum initial
condition is used, preferably.  Since Eqs.~(\ref{eq:Dx}-\ref{eq:Dy}) are first
order in time, all further evolution is determined, once the background is
known.  Of course, any other initial particle content is also feasable, one can
e.g. set an uneven number of particles and antiparticles, which is the
microcanonical analog of a baryochemical potential. We will give formulas where
these particle numbers enter later below.

\subsection{\label{sec:mfe}Mode function expansion}

One can solve Eqs.~(\ref{eq:Dx}-\ref{eq:Dy}) and (\ref{eq:Phi}) numerically
without any further information. The standard strategy is to introduce
mode functions, i.e. to treat time evolution as a Bogolyubov transformaton
of the initial-time ladder operator. This method has been formerly used for
bosonic fluctuations on a homogeneous background
\cite{AlamosLO,ParisLO,BaackeON}, and later extended to fermionic systems
\cite{BaackeFerm,FermionicPreheating} and also to inhomogeneous backgrounds
\cite{AmsterdamInhom,BettencourtInhom}. The equations
for fermionic fluctuations on an inhomogeneous backgrounds have been worked out
in detail by Aarts and Smit \cite{AartsSmitNPB}.

We introduce the mode functions $\phiu$ and $\phiv$ as classical solutions
weighting the anticommuting ladder operators with
\begin{eqnarray}
&&\anticomm{b_s(\p)}{b_{s'}^+(\q)}=(2\pi)^3\delta(\p-\q)\delta_{s,s'}\,,\quad\\
&&\anticomm{d_s(\p)}{d_{s'}^+(\q)}=(2\pi)^3\delta(\p-\q)\delta_{s,s'}\,\quad
\end{eqnarray}
in the fermion field operator:
\begin{equation}
\Psi(x)=\intp\sum_s\left( b_s(\p)\phiu + d^+_s(\p)\phivm\right)
\label{eq:mfePsi}
\end{equation}
We introduced the spinor index $s$ that runs from 1 to 2.
If the fermions' initial condition is homogeneous, one has
\begin{eqnarray}
\evalat{\phiu}{x^0=0}&=&u^s(\p)e^{-ip_jx^j}\,,\\
\evalat{\phiv}{x^0=0}&=&v^s(\p)e^{-ip_jx^j}\,.
\label{eq:mfeic}
\end{eqnarray}
The ladder operators correspond to these initial time excitations that are
transformed as fermions travel through the background.  The statistical
features of these operators actually reflect the initial particle distribution:
\begin{eqnarray}
&&\avr{\comm{b^s(\p)}{{b^{s'}}^+(\q)}}
=(2\pi)^3\delta(\p-\q)\delta_{s,s'}(1-2 n^s_+(\p))\,,\qquad\\
&&\avr{\comm{d^s(\p)}{{d^{s'}}^+(\q)}}
=(2\pi)^3\delta(\p-\q)\delta_{s,s'}(1-2n^s_-(\p))\,.\qquad
\end{eqnarray}
The $u^s(\p)$ and $v^s(\p)$ spinors in Eq.~(\ref{eq:mfeic}) are defined as the
eigenvectors of the vacuum correlation matrix written momentum space:
\begin{equation}
{\cal M}(\p)=\frac{1}{\omega_\p}\left(p_i\gamma^i\gamma^0+m\gamma^0\right)\,.
\label{eq:M}
\end{equation}
This matrix has the eigenvalues $(+1,+1,-1,-1)$ corresponding to the
eigenvectors $u^1(\p)$, $u^2(\p)$, $v^1(\p)$ and $v^2(\p)$ respectively.
On a non-trivial background these eigenvalues disambiguate between particle
and antiparticle solutions. Using the identities
\begin{eqnarray}
\gamma^0v^s(-\p)&=&v^s(\p)\,,\\
\sum_s\left( u^s(\p){u^s}^+(\p)+v^s(\p){v^s}^+(\p)\right)&=&1\,,\\
\sum_s\left( u^s(\p){u^s}^+(\p)-v^s(\p){v^s}^+(\p)\right)&=&{\cal M}{\p}
\end{eqnarray}
one can show that at initial time the two-point function in Eq.~(\ref{eq:Dic}) is
correctly reproduced by the field operator in Eq.~(\ref{eq:mfePsi}).

At any later $x_0$ the mode functions are given by the following commutators:
\begin{eqnarray}
\avr{\comm{\Psi(x)}{{b^s}^+(\p)}}&=&\phiu\,,\\
\avr{\comm{\Psi(x)}{{d^s}(-\p)}}&=&-\phiv\,.
\end{eqnarray}
On the other hand, one can express the ladder operators in terms of the
initial time field operator by
\begin{eqnarray}
{b^s}^+(\p)&=&\int_{\x} \evalat{\Psi^+(x)}{x_0=0} u^s(\p)e^{-i\p\x}\,,\\ 
{d^s}^(-\p)&=&\int_{\x} \evalat{\Psi^+(x)}{x_0=0} v^s(\p)e^{-i\p\x}\,.
\end{eqnarray}
Using these one has
\begin{eqnarray}
\phiu=2\int_\p e^{-i\p\y}\evalat{D(x,y)}{y_0=0} \gamma^0 u^s(\p)\,,\\
\phiv=-2\int_\p e^{-i\p\y}\evalat{D(x,y)}{y_0=0} \gamma^0 v^s(\p)\,.
\label{eq:mfephifromavr}
\end{eqnarray}
These equations relate the propagators used in
Eqs.~(\ref{eq:Dx}-\ref{eq:Dy}) to the mode functions.
So that $\Psi(x)$ in Eq.~(\ref{eq:mfePsi}) solves the Dirac
equation~(\ref{eq:Dirac}) the mode functions $\phiu$ as well as $\phiv$
have to solve the same Dirac equation for all $\p$ and $s$. 
\begin{equation}
(i\gamma^{\mu}\partial_{\mu}-m+g\Re\Phi(x)-ig\Im\Phi(x)\gamma^5)\phi^{u/v,s}(x,\vec p)=0 \,,
\label{eq:mfedirac}
\end{equation}
This is now also manifest from Eq.~(\ref{eq:mfephifromavr}).
We can actually confirm the initial condition in Eq.~(\ref{eq:mfeic}) by
inserting $D(x,y)$ of Eq.~(\ref{eq:Dic}) into Eq.~(\ref{eq:mfephifromavr}).

\subsection{\label{sec:renormalisation}Renormalisation}

The effective potential in Eq.~(\ref{eq:effpot}) has a non-polynomial
contribution from the logarithm of the fermion determinant. Expanding
in $\Phi$ to $n$-th order one finds the fermion one-loop diagrams with $n$
external bosonic lines.  These diagrams with $n\le4$ are potentially divergent
in 3+1 dimensions. 
Already at $n=1$, the source (\ref{eq:J}) is quadratically divergent.

We renormalise the scalar potential additively by introducing
a renormalised potential $V$ and a counterfunction $\delta V'(\Phi)$
in Eq.~(\ref{eq:Phi}). We also introduce a wave function renormalisation
so that the renormalised scalar evolution equation reads
\begin{equation}
Z\partial^2_x\Phi_R(x)+V_R'(\Phi_R(x))+\delta V'(\Phi_R(x))+N_f J(x)=0\,.
\label{eq:renphi}
\end{equation}
To calculate $\delta Z=Z-1$ we linearise $J$ in $\Phi$ and obtain
\begin{eqnarray}
&&Z\partial^2_x\Phi_R(x)+V_R'(\Phi_R(x))+\delta V'(\Phi_R(x))=\nn
&&\qquad N_f \int_0^{x_0} dz_0\int d^3 z \Sigma(x-z)\Phi_R(z)\,.
\label{eq:linearphi}
\end{eqnarray}
Here $\Sigma(x)$ stands for the vacuum one-loop self energy.  The counterterms
$\delta Z$ and $\delta \mu^2$ (see Eq.~(\ref{eq:massct}) below)
will be set so that they cancel the potentially divergent first two coefficients
in the $k^2$  expansion of $\Sigma(k_0,\vec k)$ so that the renormalised
self energy
\begin{equation}
\Sigma_R=\delta Z k^2 +\delta m^2 +\Sigma(k_0,\vec k)
\label{eq:renSigma}
\end{equation}
is finite in the perturbative vacuum of the fermions.

Following
the existing practice in classical simulations, we will use a temporal
discretisation step that is negligible to the spatial lattice spacing,
i.e the cut-off is three-dimensional, and the three-dimensional momentum
integrals are implicitly regularised. We give an explicit form of $\Sigma$
in the spatial Fourier space:
\begin{equation}
\Sigma(t,\vec k)=-4g^2\intp \left[
\frac{m^2-\vec p(\vec p-\vec k)}{{\omegap}\omega_{\vec k-\vec p}}
-1\right]
\sin{\omegap}t\,\cos\omega_{\vec k-\vec p}t\,.
\label{eq:Sigma}
\end{equation}
We define ${\omegap}=\sqrt{m^2+p^2}$.
The wave function renormalisation we either get by taking the second
$k$-derivative at scale of renormalisation, which is $k=0$ in our calculation,
or one calculates it from the real-time behaviour using the formula
\begin{equation}
\delta Z=N_f\int_0^\infty dt \frac{t^2}{2} \left.\Sigma(t,\vec k)\right|_{\vec k=0}\,.
\label{eq:dZintegral}
\end{equation}
This equation is the real-time variant of
$\delta Z=N_f\partial^2\Sigma(k_0,\vec k)/(\partial k_0)^2$ at zero momentum
and makes sure that the coefficient of $k^2$ vanishes in
Eq.~(\ref{eq:renSigma}).

One can perform the time integral in Eq.~(\ref{eq:dZintegral})
under the assumption that oscillations
of the indeterminate integral at large times are incoherent and they are
averaged away when the $\vec k$-integral is carried out. One finally
arrives at
\begin{equation}
\delta Z=-\frac{N_fg^2}{2}\intp
\frac{p^2}{{\omegap}^5}\,,
\end{equation}
the divergence is logarithmic, as expected. An analogous calculation
delivers the scalar mass counterterm
\begin{equation}
\delta \mu^2=\int_0^\infty dt 
\left.\Sigma(t,\vec k)\right|_{\vec k=0}=
2N_fg^2\intp
\frac{p^2}{{\omegap}^3}\,,
\label{eq:massct}
\end{equation}
which is quadratically divergent.

To renormalise the coupling we need to go beyond the linear approximation in
Eq.~(\ref{eq:linearphi}). We renormalise the effective potential at zero
momentum. We analyse the non-linear response to a static field and compensate
the force on this static field by $\delta V'$. This way we do more than
substracting divergences. We actually alter the finite part of the theory
so that the scalar potential is exactly as it was before coupling to fermions.
This complete renormalisation will ensure the correctness of any comparison
with the purely bosonic classical field theory.

A static scalar field with a Yukawa coupling is similar to a Dirac mass.
The current $J$ is then constant in space and time, but it depends on the
mass $M=m-g\Phi_R$. The counterterm $\delta V'(\Phi_R)$ based on the vacuum
one-loop diagrams reads
\begin{equation}
\delta V'(\Phi_R)= -2 N_fg \intp
\frac{m-g\Phi_R}{\sqrt{(m-g\Phi_R)^2+p^2}}
\label{eq:deltaVprime}
\end{equation}
Expanding this integral to linear order in $\Phi$ gives the same counterterm
as we have already found in Eq.~(\ref{eq:massct}). To third order in $\Phi$
we find in the chiral limit for the coupling renormalisation
$\delta\lambda=12\delta Z$ as it has been also derived in \cite{BaackeFerm}.

Of course, the integral in Eq.~(\ref{eq:deltaVprime}) would be very
time consuming to calculate in each space-time point when solving
Eq.~(\ref{eq:renphi}). Therefore we approximate $\delta V'$ with a fifteenth
order polynomial fitted in the range $ag\Phi\in [-2.5,2.5]$. The relative
precision of the fit is between 1 and 10\%,
(the greatest when $\Phi_R\approx 0$).
The fit interval is exceeded only by extreme excitations on coarse lattices,
and one can extend it with little effort.

In the rest of the paper we do not write out the $R$ index for the
renormalised background, and all parameters are understood as renormalised.
For simplicity, we also hide the counterterms in the equations we discuss,
but we keep them in our numerics, of course.

Contrary to Ref.~\cite{BaackeFerm}, in this approach we
solve equations with divergences, which cancel in the end result. This makes
the final removal of the cut-off impossible, and such a calculation is usually
error-prone close to the continuum limit. But in this case we solve a lattice
field theory classically and it makes no sense to even approach the continuum
limit. This renormalisation makes sure that the fermionic vacuum does not
alter the bosonic vacuum, but the classical divergences from the closed
bosonic loops are as dangerous as before. 

By construction, fermions have now no impact on a static bosonic field,
but there is a damping rate for dynamical fields, which is given in
the real scalar case by
\begin{equation}
\gamma(k)=\frac{1}{2k_0}\int_0^\infty dt \sin(k_0 t) \Sigma(t,\vec k)\,.
\end{equation}
For a scalar with mass $\mu$ this evaluates for homogeneous mode to
\begin{equation}
\left.\gamma(\mu,\vec k)\right|_{\vec k=0}=\frac{g^2N_f\pi}{2\mu}
\intp\delta(\mu/2-\omega_p)
=\frac{g^2N_f}{16\pi\mu^2}(\mu^2-4m^2)^{3/2}\,,
\end{equation}
if $\mu>2m$. The damping rate is directly observable from the numerics.
Since it is proportional to $N_f$, it facilitates the measurement
of the number of doublers in a lattice implementation.

\section{\label{sec:stochastic}Stochastic approach}

In Eq.~(\ref{eq:mfedirac}) of the previous section a separate field has to be
evolved for each mode $\vec p$ and spinor index. On a three-dimensional lattice
with $N^3$ sites this means a coupled set of $4\cdot 4\cdot N^6$ complex
ordinary differential equations. This relatively high price might explain the
fact that in nearly ten years time since the equations have been published no
calculation has been carried out beyond 1+1 dimensions.

An elegant way of performing integrals with high dimensionality 
is to employ Monte-Carlo techniques. Importance sampling is a prominent example
in statistical field theory, though its formulation for fermionic 
fields is troublesome because of the Grassmann nature of these degrees of
freedom. Nevertheless, there have been promising news to the apparently
impossible simulations at finite chemical potential \cite{AartsStochastic} or
in real time \cite{BergesStochastic}.

In fact, the situation in  our semiclassical nonequilibrium setting is much
simpler than in Euclidean field theory simulations. We know everything about the
initial fermion ensemble and we will set up evolution equations for the members
of this ensemble. At any later time an averaging over these members will tell
the propagator $D(x,y)$.

Notice that we could formulate Eqs.~(\ref{eq:Dx}-\ref{eq:Dy}) as well
as the back reaction (\ref{eq:J}-\ref{eq:JPS}) in Eq.~(\ref{eq:Phi}) without
any reference to the spectral function, which is complementary to the
symmetrised propagator $D(x,y)$. We will replace the commutator of
anticommuting operators by the product of plain complex numbers in $D$.
To accept this simplification we have to show that the two-point function
defined in terms of this simple product obeys the same equations of motion as
$D$ and that it also starts from the same initial condition.

Let us introduce a set of classical spinor stochastic variables as
$c$-number fields:
$\psi_M(x)$ and $\psi_F(x)$. Only together can these ``male'' and ``female''
fields form a meaningful physical quantity, but the male and female roles
are interchangeable:
\begin{equation}
D(x,y)=\avr{\psi_M(x)\bar\psi_F(y)}=\avr{\psi_F(x)\bar\psi_M(y)}\,.
\label{eq:Dconst}
\end{equation}
The reason for why we need two spinor fields is that with a single spinor
field only positive semidefinite correlators can be modelled, whereas ${\cal
M}(\vec p)$ in Eq.~(\ref{eq:M}) has negative eigenvalues.  

So that $D$ in Eq.~(\ref{eq:Dconst}) obeys Eqs.~(\ref{eq:Dx}) and (\ref{eq:Dy})
we require that both the male and female stochastic spinors follow the
usual Dirac equation:
\begin{equation}
(i\gamma^{\mu}\partial_{\mu}-m+g\Re\Phi(x)-ig\Im\Phi(x)\gamma^5)\psi_g(x)=0 \,.
\label{eq:psi}
\end{equation}
The $g$ (gender) index represents $M$ or $F$.

The currents expressed in terms of the stochastic fields read
\begin{eqnarray}
J^{\rm S}(x)&=&g\Tr D(x,x) = g \avr{\psi_F^+(x)\gamma^0\psi_M(x)}\,,\\
J^{\rm PS}(x)&=&g\Tr D(x,x)\gamma^5 = g \avr{\psi_F^+(x)\gamma^0\gamma^5\psi_M(x)}\,.
\end{eqnarray}
Due to the interchangeability of $\psi_M$ and $\psi_F$ the scalar and
pseudoscalar currents are manifestly real and imaginary, respectively. 

We have to make sure to satisfy Eq.~(\ref{eq:Dic}). For this we define the
Fourier transformed stochastic fields:
\begin{equation}
\psi_{g}(\vec p)=\int_\x e^{ip_jx^j}\psi_{g}(\vec x),
\quad
\psi_{g}(\vec x)=\int_\p e^{-ip_jx^j}\bar\psi_{g}(\vec p)\,.
\end{equation}

To reproduce Eq.~(\ref{eq:Dic}) we require
\begin{equation}
\avr{\psi_M(\p)\psi^+_F(\q)}=(2\pi)^3\delta(\p-\q)\frac{1}{2}{\cal M}(\vec p)
\label{eq:pic}
\end{equation}
To actually realise an initial ensemble with this correlator one has to
solve the eigenvalue problem of ${\cal M}(\vec p)$. This we have actually
done already when we introduced the mode functions and denoted the eigenspinors
as $u^{(1)}$ ,$u^{(2)}$, $v^{(1)}$ and $v^{(2)}$ corresponding to the
eigenvalues $+1,+1,-1$ and $-1$ respectively.

We can express the stochastic spinor fields
in terms of the eigenspinors as follows:
\begin{equation}
\psi_{M,F}(\vec p)=
\frac1{\sqrt{2}}\sum_{s}\left(
\xi_s(\p)u^s(\p)
\pm
\eta_s(\p)v^s(\p)
\right)
\label{eq:psimfinit}
\end{equation}
$\xi^{s}$ and $\eta^{s}$ are the primary complex random variables we use:
\begin{eqnarray}
\avr{\xi^s(\p){\xi^{s'}(\p)}^+}&=&(2\pi)^3\delta(\p-\q)\delta_{s,s'}(1-2n^s_+(\p))\,,\nonumber\\
\avr{\eta^s(\p){\eta^{s'}(\p)}^+}&=&(2\pi)^3\delta(\p-\q)\delta_{s,s'}(1-2n^s_-(\p))\,.\nonumber
\end{eqnarray}

All other two-point correlators vanish. (Actually, these variables
could be chosen real and do not necessarily
have to be Gaussian.) Notice that nothing on the right hand side of
Eq.~(\ref{eq:psimfinit}) bears a gender index, but the male and female fields
have different signs for the antiparticle component. This allows for the
stochastic representation of the hermitian matrix with negative eigenvalues in
Eq.~(\ref{eq:pic}). With $\xi$ and $\eta$ we actually simulate the ladder
operators: this is possible since the ladder operators always appear in
the expectation value of a commutator. 

The eigenvalues of the correlator $\avr{\psi_M(\p)\psi_F^+(\p)}$, which is a
matrix in Dirac indices, actually represent the particle number: they take the
vaule $\frac12-n_+^{(s)}(\p)$ for the fermions, and $n_-^{(s)}(\p)-\frac12$ for
the antifermions. By proper initialisation, one can start from a polarised
fermion gas, or, one can set a constant non-vanishing baryon density, as we
anticipated.  In a completely symmetric setting we can read out the particle
number by taking the determinant of the correlation matrix (in momentum space),
which shall be $\left(n(\p)-\frac12\right)^4$.

At this point we return to the question of numerical feasibility. The
expectation value in Eq.~(\ref{eq:Dconst}) turns into an average over $E$ pairs
of spinor fields in practice, where $E$ is finite number.  The statistical
error in Eq.~(\ref{eq:Dconst}) propagates through
Eqs.~(\ref{eq:J}-\ref{eq:JPS}) into the scalar equation. The statistical noise
in the back reaction may induce artifical production of scalar fluctuations.
Thus, checking for the $E$-dependence of the final result is an essential part
of using this scheme. If the required number of spinor pairs ($E$) turns out to
be higher than the number of lattice sites, the standard deterministic mode
function expansion is the cheaper and more precise option. This is typically
the case in 1+1 dimensions. Increasing the number of dimensions, however, one
can in most cases keep $E$ around the linear lattice size or less, and the
stochastic method can be by several orders of magnitude more efficent than the
deterministic algorithm, both in memory need and in time.

For future reference we give the actual form of the spinor equations as well
as their initialisation in Appendix \ref{sec:spinors}. Since the system
we analyise is implicitly understood to be discretised on a lattice, some
comments on lattice doublers are due in Appendix \ref{sec:doublers}.


\section{\label{sec:therm}Effective scalar dynamics}

In this section we present the numerical analysis of a real scalar field
coupled to fermions as introduced above. For the sake of simplicity of the
implementation we restrict our numerics to 2+1 dimensions.

\begin{figure}
\centerline{\includegraphics[width=3.5in]{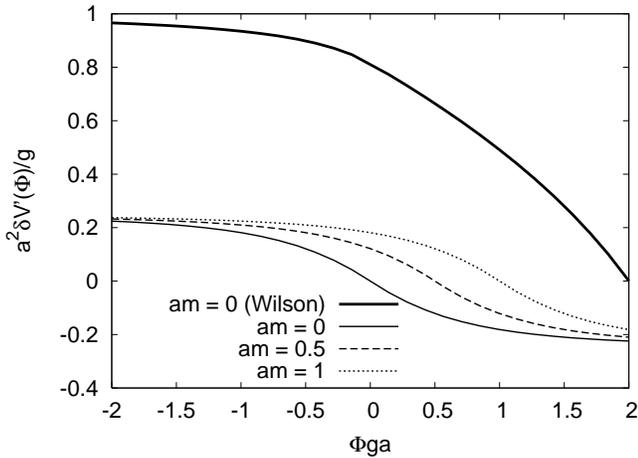}}
\caption{\label{fig:deltaV}
Renormalisation of the effective potential. For the thick line we used
Wilson fermions in two spatial dimensions with $r=1$. The other lines
have been calculated in the presense of doublers.  The breaking
of chiral symmetry manifests in the asymmetry of $\delta V'(\Phi)$ around zero.
}
\end{figure}

We perform the renormalisation of the effective potential as already
anticipated, but no wave function renormalisation is necessary. In
Fig.~\ref{fig:deltaV} we give $\delta V'(\Phi)$
by evaluating the two dimensinoal variant of Eq.~(\ref{eq:deltaVprime})
on a large lattice for various fermion masses.
In the plot we used $a$ for the lattice spacing.  Notice that in the massive
case with broken chiral symmetry we loose the $\Phi\leftrightarrow-\Phi$
symmetry. For this reason we use massless fermions and compensate for
the doublers as detailed in Appendix \ref{sec:doublers}.

In the following we discuss a few test cases to explore the capabilities
of this semiclassical approximation. To better see the effects of the fermions
we always run the purely classical simulation with the same initial condition
(same random seed) in parallel. For reproducibility we give the
parameters in the figure captions: the linear lattice size ($N$), the
Yukawa coupling ($g$), the fermion's inital temperature ($T_f$), the scalar mass
($\mu$) and coupling ($\lambda$), and the number of spinor fields ($E$) in the
ensemble. In these experiments we used two-component chiral
fermions. These parameters and the data on the plots are given in lattice units
(with $a=1$).

\begin{figure}
\centerline{\includegraphics[width=3.5in]{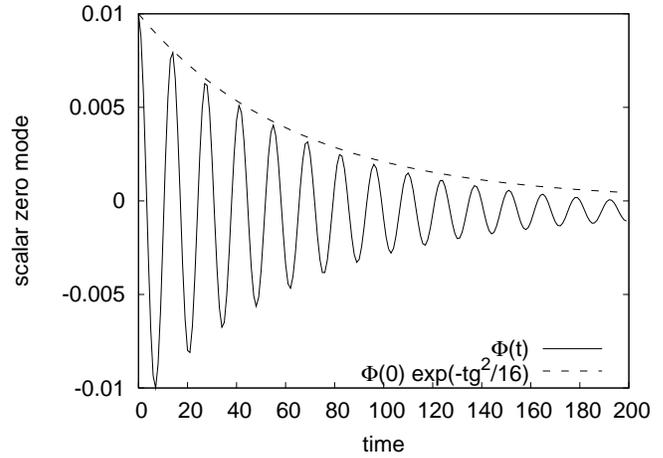}}
\caption{\label{fig:damping}
The homogeneous mode of the scalar field is exponentially damped at the
expected rate.
(Parameters: $N=1024$, $g=0.5$, $T_f=0$, $\mu^2=0.25$, $\lambda=0$ and
$E=20$)
}
\end{figure}

In our first exercise we plot the damping of the scalar homogeneous mode in
Fig.~\ref{fig:damping}. The exponential with expected rate ($\gamma=g^2/16$ in
2+1 dimensions with 2-component spinors) nicely forms an envelope of the
calculated
evolution. It was important to use a large volume, otherwise the damping
stopped at about $N/2$ time and recurrences occur. In fact, one assumes
infinite volume in the derivation of the decay rate. 

Let us now consider an example where the fermions start from a finite
temperature state and transfer energy to the bosonic vacuum. We set 
up an experiment with a small noise in the bosonic sector, $\mu^2=0.25$,
$\lambda=6$, $g=0.25$ and $T_f=1$. To our surprise, there was no boson
production at all, but the small initial scalar noise was transformed
into fermions with a rate comparable to $\gamma$. It seems that
in the semiclassical approximation the production of quantum fluctuation
is a one-way channel of interaction.

\begin{figure}
\centerline{\includegraphics[width=3.5in]{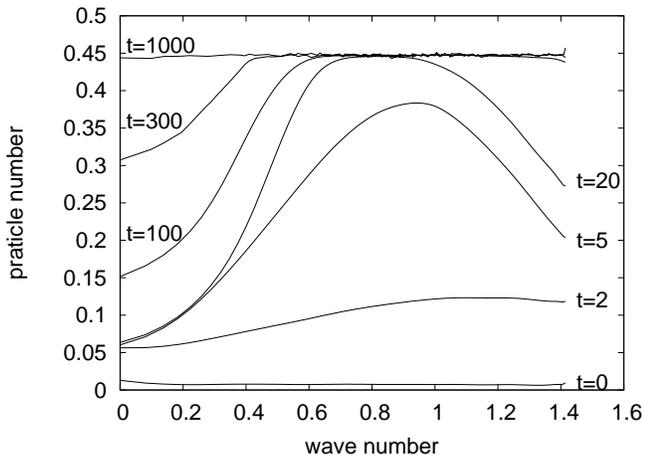}}
\caption{\label{fig:therm5sp}
Anti-thermalisation. The bosonic excitations are transformed to fermions
until fermion-production is cut by Pauli blocking and a close-to-infinity
temperature sets in.  
(Parameters: $N=64$, $g=0.5$, $T_f=0$, $\mu^2=0.25$,
$\lambda=24$ and
$E=32$, average of 20 runs.)
}
\end{figure}

To see how the energy is transferred to fermions regardless to
our thermodynamical preconceptions we present the results of our third
experiment. The scalar field is now started from a non-thermally excited
state with an isotropic particle distribution peaked around the momenta
$|\vec k_0|= 0.5$ with $n(|\vec k_0|)=10$. The initial energy density was
$\approx1$. What we see in Fig.~\ref{fig:therm5sp} is a counter-intuitive
anti-thermalisation, where all energy that can be possibly transformed
to quantum fluctuations is taken away from the background. This also
happens in the purely bosonic Hartree approximation, but here in the fermionic
case the particle number is capped at a value close to 1/2 due to Pauli
blocking.  The modes above $|\vec k|>1$ are quickly excited (at the order of
damping time).  The low momentum modes are filled up on a much slower scale.
At this point we remark that a two-dimensional classical scalar theory
comes into non-thermal (quasi) fixed points for a wide range of initial
conditions. For a similar classical system we found that the evolution
to equilibrium can be extremely slow, governed by a power law
\cite{BorsanyiO2}. In this example, too, the scalar spectrum evolves
into an approximate power law with an exponent of $\approx-1.8(2)$. The effects
of fermions manifests merely as an overall coefficient in the spectrum.

\begin{figure}
\includegraphics[width=3.5in]{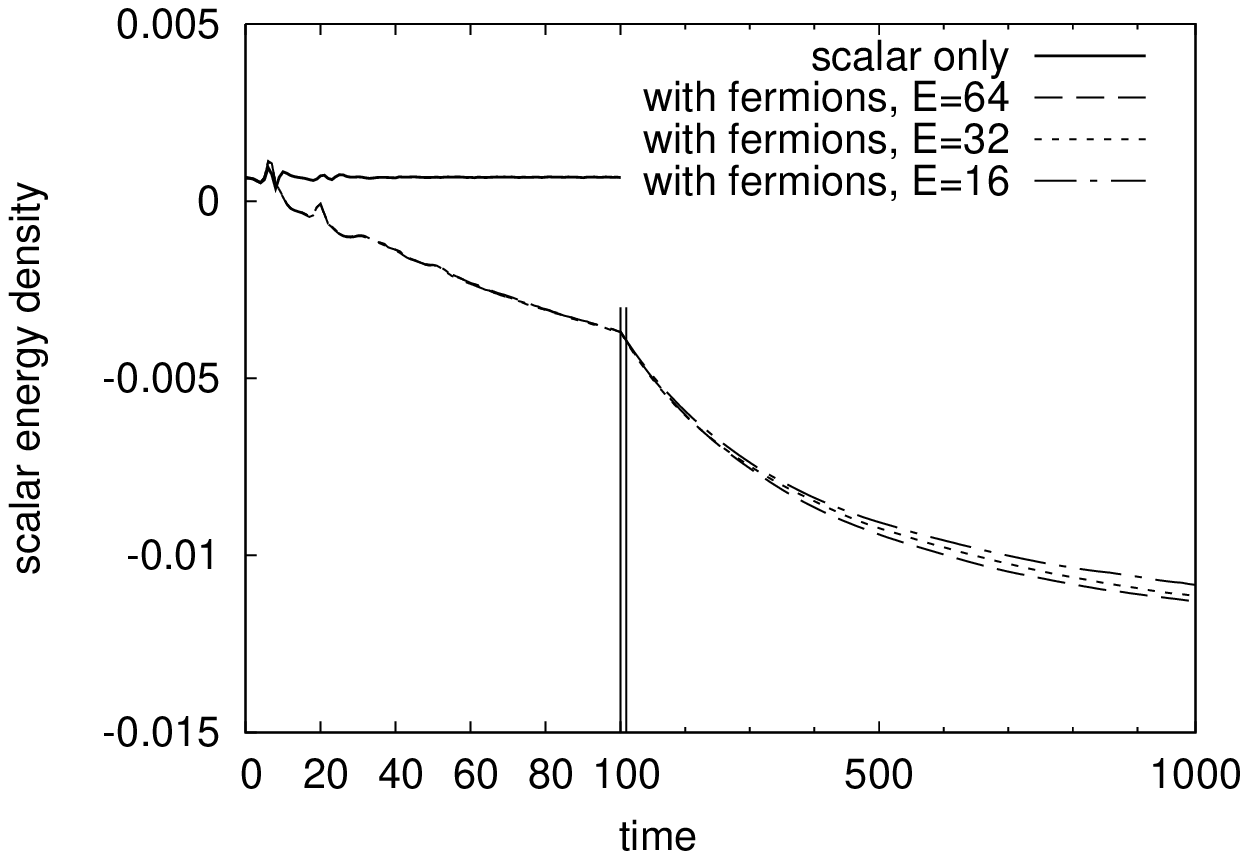}

\includegraphics[width=3.5in]{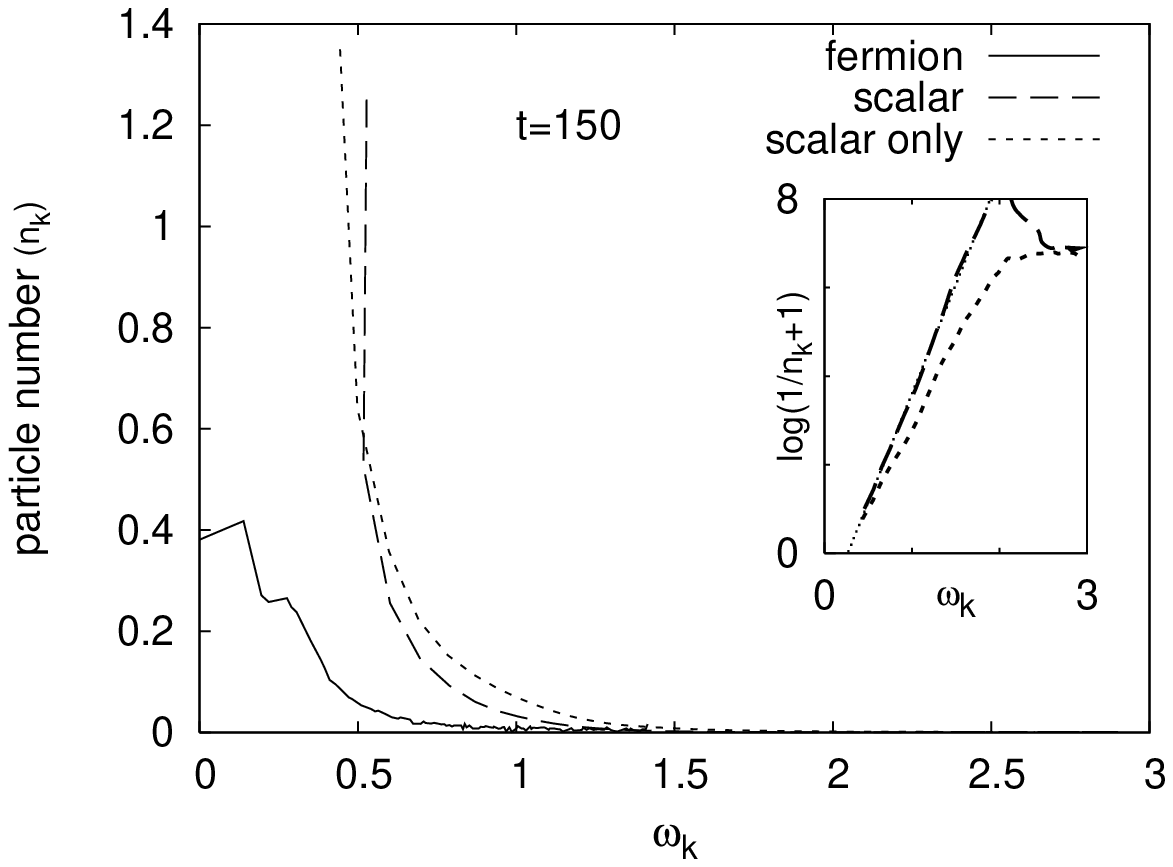}
\caption{\label{fig:therm4}
Scalar field with a spinodal instability. 
{\it Top:} energy density of the scalar field with or without coupling to
fermions. For comparison, three different
{\it Bottom:} the particle spectra at $t=150$. 
At and around the time shown, the scalar spectrum is close the
Bose-Einstein distribution, especially when coupled to fermions. The dotted
line in the inset plot is the thermal fit ($\beta=4.9$).
(Parameters: $N=64$, $g=0.25$, $T_f=0$, $\mu^2=-0.25$,
$\lambda=6$ and $E=64,32,16$, average of 20 runs.)
}
\end{figure}

We finally show an example where the classical approximation is expected
to work well. We start the classical system from the center of a double-well
potential. There is a rapid particle production fueled by the spinodal
instability. The resulting scalar spectrum is not far from a Bose-Einstein
distribution. Of course, this closeness to quantum equilibrium is temporary:
the slow classical thermalisation drives the system towards classical equipartition.
Coupling this scalar field to fermions switches on a dissipation, as
one can see in the plotted energy density in Fig.~\ref{fig:therm4}.

The scalar spectra in Fig.~\ref{fig:therm4} are close to a quantum
equilibrium distribution with some chemical potential. The UV end of the
spectrum is distorted by lattice artefacts, but otherwise the linear fit
of $\log(1/n(\vec p)+1)$ is adequate
(see dotted line in Fig.~\ref{fig:therm4}). The closeness to the
quantum equilibrium distribution is maintained throughout the time we followed
the dynamics. The energy drain of the fermion field does not bring the
scalars out of this equilibrium but imposes a steady cooling.

These numerical experiments lead us to the negative conclusion that the energy
transfer between the classical and quantum degrees of freedom is
unidirectional. As
it is also known, the produced quantum particles do not scatter on each
other, and the energy transfer between modes through the inhomogeneous
background is inefficient. Can this semiclassical approximation be used then
at all?

We do think that in some circumstances this low-cost solution to add fermions
to a classical field simulation is adequate. The rate of fermion production
is correctly given account for, albeit these particles will not thermalise.
The very mechanism of particle production and the simultaneous loss of
energy in the bosonic sector is well described as long as fermions are
not created in such an abundance that their non-thermal distribution could have
impact on the back-reaction. In fact, bulk observables, such as the scalar
effective potential, prethermalise \cite{Prethermalization}, i.e their
value before thermalisation can be used as an estimate to what one would find
after equilibration. Even if the fermion distribution is non-physical,
the evolution of the bosonic background can be well approximated. If, however,
the back-scattering of the produced fermions to bosons becomes relevant,
this semiclassical approximation will no longer be applicable. One can
actually monitor the fermion particle numbers to check for relevance of
(the absence of) back-scattering. 


\section{\label{sec:oscillon}Fermionic decay of oscillons}

We can consider the semiclassical approximation safe if the resulting fermion
energy density is small. However, one of the justifications for the classical
approximation is the high bosonic occupancy, which will inevitably generate
an energy transfer into the fermionic fields. In such cases the approximation
will break down within a short time, which is likely to be the damping time
we discussed in the previous section. 

In the this section we turn to applications where classicality has an
other justification. If the particle content is very low and but there is
dilute network of classical structures, such as topologial defects,
their evolution can be well described by the non-linear wave equations.
The decay of these structures into particles is mapped to the production of
classical waves (``ripples'') by the classical equations. As this mostly
happens in the ultraviolet, the classical approach is not justified for
describing particle production, in contrast to its usefulness in the case
of the macroscopic networks, like cosmic strings.

We addressed this deficiency of the classical approximation in 
Ref.~\cite{SemiclassicalDefects}, where we introduced a stochastic approach
to the bosonic mean-field approximation, similar to the method presented in
this paper. We found that mimicking the quantum distribution by an analogous
classical noise (following the so called ``just-the-half'' prescription)
introduces undesired time-dependent renormalisation effects to the effective
potential. Instead we solved the inhomogeneous mean-field equations and found
that on the macroscopic level, the decay channel into quantum particles
plays negligible role, whereas it on microscopic scale we found deviations.
Our numerical analysis suggested that oscillons, which are one of the classical
decay products of topological defects \cite{SussexOscillon}, are the primary
sources of quantum particles, while the direct radiative decay of a defect
network is suppressed as predicted in Ref.~\cite{SrednickiTheisen}.

To better understand how oscillons decay quantum mechanically we reproduce
one of the experiments in Ref.~\cite{SussexOscillon}, but we also add fermions.
In two dimensions oscillons are particularly stable
\cite{SaffinTranbergOscillon,GleiserAnalytic} localised structures, when
several oscillons are created in volume, they
behave as molecules in a gas. When oscillons collide, the coalesce with some
probability. Being this mechanism their only decay process (in 2+1 dimensions),
the number density of oscillons obeys the equation $\dot n(t)\sim n^2(t)$.
Thus, the classical solution is $n(t)\sim 1/t$, which is approximately manifest
in classical simulations \cite{SussexOscillon}.

\begin{figure}
\centerline{\includegraphics[width=3.5in]{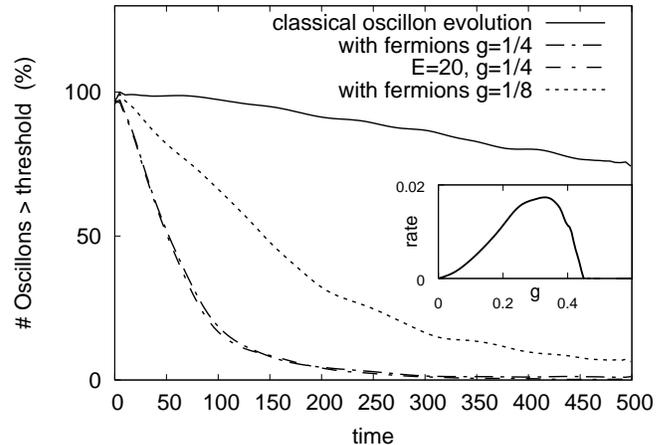}}
\caption{\label{fig:oscillon}
Estimated ``molecule'' number in a gas of oscillons. When the scalar background
is coupled to fermions, the slow classical evolution is replaced by an
approximately exponential decay. An estimate of the rate as a function of
the Yukawa coupling is shown in the inset plot.
The obtain the same curves using the mode function expansion would have
required three orders of magnitude more computational resources.
For $g=1/4$ we explicitly check for the insensitivity to doubling the
ensemble.
(Parameters: $N=128$, $T_f=0$, $\mu^2=-0.25$,
$\lambda=3$ and $E=10$ or 20. For each coupling we averaged 30 runs. )
}
\end{figure}

We put 16 incoherent oscillons with small random velocities in a box with
$N=128$.  We estimate the number of oscillons by counting the sites with an
energy density beyond a thresold ($\varepsilon>0$). This number we normalise to
its initial value and plot in Fig.~\ref{fig:oscillon}. It takes long before the
expected power-law solution sets in (and even then finite volume effects can
distort it). But a small coupling to the fermionic fields introduces a new time
scale, and the slow classical behaviour is replaced by a
close-to-exponential decay. (The oscillon damping rate is about four times
stronger than for the homogeneous mode.) This process reduces the amplitude of
most oscillons below the threshold. After $t>100$, however, the plotted
estimate can be best fit by a power law with an exponent of $-2$.
For the semiclassical evolution of these localised objects a surprisingly small
spinor ensemble already provides
results that are insensitive to an increase in $E$.

In the inset plot we estimated the oscillon decay rate by the inverse
time necessary to radiate away $100\exp(-1)$ percent of the oscillons. The
rate cuts off at about $g_{\rm cut}\approx0.45$. One can explain this by
simple kinematics. The effective mass of the produced fermions $m_f=gv$, where
$v=\sqrt{-6\mu^2/\lambda}$ is the vev of the background. The scalar mass
in the broken phase is $m_b=\sqrt{-2\mu^2}$. It is not this bosonic
mass that enters the kinematical relation but the oscillon frequency
$\omega_{\rm osc}$, so the condition for the decay is
$\frac12\omega_{\rm osc}>m_f$. From $g_{\rm cut}$ we can tell the oscillon
frequency: $\omega_{\rm osc}/m_b=2g_{\rm cut}\sqrt{3/\lambda}\approx 0.9$.
This estimate is in harmony with direct measurements \cite{SussexOscillon}.

In this paper our aim is not to explore the parameter space, and to analyse the
mechanisms oscillon decay. Instead, we put forward a low-cost
technique to check existing and future analyses of defect evolution for fermionic 
quantum corrections, complementing work already done for bosonic 
ones\cite{SemiclassicalDefects}. We plan to investigate  
the evolution of cosmic strings for such contributions from
quantum degrees of freedom in a future publication.

\section{\label{sec:discussion}Conclusions}

In this paper we propose a low-cost integration scheme for the fermionic
path integral, which leads to equations that are equivalent to the mean-field
approximation studied earlier by Aarts and Smit. These equations also
follow from the large-$N_f$ expansion of the 2PI effecitve action.
The computational efficiency
of this scheme allowed us to do simulations beyond 1+1 dimensions. This
stochastic method is a generalisation of our earlier technique developed
for scalars in Ref.~\cite{SemiclassicalDefects}.

We calculate several test cases on a scalar example and study what the
irreversible phenomena can be captured by this simple method. We confirm
that the fermions, once created, can no longer scatter on each other, but
this is not the only obstacle that hinders thermalisation. The damping
of the classical oscillations are correctly given account for, but the
back-scattering of the fermions into bosons is absent. In the language
of the mode function expansion, fermions (and also other quantum fluctuations
on the Hartree level) are represented by far more dynamical variables that
the background. If these variables strive for classical equipartition
(as usual in a coupled set of non-linear differential equations), the energy
left in the background is negligible. This suppression of the background
becomes stronger with higher dimensionality, and was less relevant in former
1+1 dimensional calculations. 

As it was remarked in Ref.~\cite{AartsSmitPRD} the fermion spectrum can
become close to thermal, and this raised hope that the inhomogeneous Hartree
approximation is still capable to account for an approximate thermalisation.
It is, however, more likely, that it is the praticle production mechanism
that brings the fermions close to equilibrium, rather than scattering.

Even though scattering cannot drive the fermions towards equilibrium, we
expect that the back-reaction of the often not-far-from-thermal fermion field
has an approximately thermal back-reaction due to prethermalisation of
the fermionic current \cite{Prethermalization}, and the lack of thermalisation
has little impact on the background field. This assumption becomes even
more plausible if we assume that the fermions leave the scene, once created.

In some physical situations it is difficult for a fermion to leave. If we
apply the presented scheme to the Yang-Mills equations, and solve the
semiclassical chromodynamics, it will be difficult for fermions to be
reabsorbed by the plasma. This puts jet-quenching outside of the range
of validity. But a semiclassical simulation of the freeze-out of the
plasma is not ruled out by the aformentioned deficiencies.

The numerical calculation of the fermion spectrum in baryogenesis scenarios
is a more viable application. If baryogenesis is driven by a first order
phase transition, the presented equations can give account for CP violation
as well as the departure from equilibrium without relying on gradient
expansion, and thus, allowing for thin walls.  For the subsequent
thermalisation, however, one has to make further assumptions.

The scheme is best applicable for systems with low particle numbers and genuine
inhomogeneities, like a dilute network of topological defects, such as cosmic
strings.  In this context fermion production is local, and the produced
particles spred in space. This results in small praticle numbers, and we expect
that the lack of thermalisation will introduce very little distortion into the
back-reaction.

In conclusion, for cases where the inhomogeneities in the background are ment
to be ``particles'', a big volume is less relevant, and other techniques, such
as the 2PI effective action on a homogeneous ensemble may be more favourable.
For the large-scale classical simulations with inhomogeneous classical
structures, however, the inhomogeneous 2PI approach would be beyond
feasibility.  In such situations, the Hartree approximation already includes the
leading quantum corrections, as well as endorses fermions. The technique in
this paper has made these type of calculations affordable.

\begin{acknowledgments}
The authors
acknowledge the collaboration with Petja Salmi on a
related project. The numerical work has been carried out on the Archimedes
cluster of the University of Sussex. SB is funded by STFC.
\end{acknowledgments}
\appendix
\section{\label{sec:spinors}Representation of the spinor fields}

The initial conditions for the spinor fields is given in terms of the
$u^s(\p)$ and $v^s(\p)$ eigenspinors. Here we present the actual form
of these eigenvectors. 
We use the normalisation factors for a theory discretised in a volume $V$.

The formulas are based on a naive fermion action. They can, however, be
easily rewritten for the Wilson fermions by replacing
$m$ to $m+\frac12\phat^2$ when discussing the $\vec p$ mode,
with $\hat p_j=\frac{2}{a}\sin p_ja/2$.

\subsection{Chiral basis in 3+1 dimensions}

In the chiral base we define the gamma matrices as
\begin{eqnarray}
\gamma^0=\twomatrix{0&1\\1&0}\,,&
\gamma^i=\twomatrix{0&\sigma^i\\-\sigma^i&0}\,,&
\gamma^5=\twomatrix{-1&0\\0&1}\,.\nonumber
\end{eqnarray}
Consequently,
\begin{eqnarray}
\gamma^i\gamma^0=\twomatrix{\sigma^i&0\\0&-\sigma^i}\,,&
\gamma^0\gamma^5=\twomatrix{0&1\\-1&0}\,.
\end{eqnarray}

The hermitian 4-by-4 matrix in Eq.~(\ref{eq:pic}) in the chiral basis reads
\begin{equation}
\avr{\psi_M(\p)\psi^+_F(\p)}=\frac{V}{2\omega_\p}
\left(\begin{array}{cc}
\vec p\vec\sigma&m\\
m&-\vec p\vec\sigma
\end{array}\right)
\end{equation}

Here $\vec p\vec\sigma$ stands for the combination of the Pauli matrices:
\begin{equation}
\vec p\vec \sigma=\twomatrix{\bar p_3&\bar p_1-i\bar p_2\\
\bar p_1+i\bar p_2&-\bar p_3}
\end{equation}
We used the standard notation $\bar p_j=a \sin p_ja$.

In the chiral base the eigenvectors are given as
\begin{equation}
\begin{array}{cc}
u^{(1)}=\alpha 
\left(\begin{array}{c}
\pabs+\omega_\p\\
m
\end{array}\right) \otimes \chi^+_p
&
u^{(2)}=\alpha 
\left(\begin{array}{c}
m\\
\pabs+\omega_\p
\end{array}\right) \otimes \chi^-_p\\
v^{(1)}=\alpha 
\left(\begin{array}{c}
-m\\
\pabs+\omega_\p
\end{array}\right) \otimes \chi^+_p
&
v^{(2)}=\alpha 
\left(\begin{array}{c}
\pabs+\omega_\p\\
-m
\end{array}\right) \otimes\chi^-_p
\end{array}
\end{equation}
with $\alpha^2=\frac{1}{2\omega_\p(\omega_\p+\pabs)}$. $\chi^\pm(\p)$ 
denote the eigenvectors of $p_i\sigma^i$ for the eigenvalues $|\vec p|$ and
$-|\vec p|$, respectively.

Let us now diagonalise $p_j\sigma^j$:
\begin{eqnarray}
\chi^+_p=\beta_+\left(\begin{array}{c}
p_3+|\p|\\
p_1+ip_2
\end{array}\right)
&
\chi^-_p=\beta_+\left(\begin{array}{c}
-p_1+ip_2\\
p_3+|\p|
\end{array}\right)
&
\textrm{if $p_3>0$ }\nonumber\\
\chi^+_p=\beta_-\left(\begin{array}{c}
p_1-ip_2\\
|\p|-p_3
\end{array}\right)
&
\chi^-_p=\beta_-\left(\begin{array}{c}
p_3-|\p|\\
p_1+ip_2
\end{array}\right)
&
\textrm{if $p_3<0$ }\nonumber
\end{eqnarray}
with $\beta_\pm^2=1/2|\p|(|\p|\pm p_3)$. The two cases we handle separately
for numerical stability (e.g. to avoid divisions by zero).

We actually solve the Dirac equation (\ref{eq:psi}) for the $\psi_g$ field
instances. This reads in chiral base
\begin{equation}
\partial_0\psi_g(x)=
\left(
\begin{array}{cccc}
\partial_3&\partial_1-i\partial_2&-iM_x^*&0\\
\partial_1+i\partial_2&-\partial_3&0&-iM_x^*\\
-iM_x&0&-\partial_3&-\partial_1+i\partial_2\\
0&-iM_x&-\partial_1-i\partial_2&\partial_3
\end{array}
\right)
\psi_g(x)\,
\end{equation}
with $M_x=m-g\Phi(x)$. We can use this equation for both $\psi_M$ and $\psi_F$.
For Wilson fermions $M_x \to M_x-\frac12\triangle$.

\subsection{Majorana basis in 2+1 dimensions}

Here we work out the implementation details for a 2+1 dimensional
setting with a real scalar background. In 2+1 dimensions we have
2-by-2 gamma matrices. In Majorana basis all these are imaginary:
\begin{eqnarray}
\gamma^0=\twomatrix{0&-i\\i&0},&
\gamma^1=\twomatrix{i&0\\0&-i},&
\gamma^2=\twomatrix{0&i\\i&0}\,.\nonumber
\end{eqnarray}
In 2+1 dimensions one of the gamma matries can be easily expressed by others:
\begin{equation}
\gamma^0\gamma^1=i\gamma^2,\qquad
\gamma^0\gamma^2=-i\gamma^1\,.
\end{equation}
The Dirac equation in this basis (with $M_x=m-g\Phi$) reads:
\begin{equation}
\partial_0\psi=\twomatrix{-\partial_2&\partial_1-M_x\\
\partial_1+M_x&\partial_2}\psi\,.
\end{equation}
The advantage of the Majorana basis becomes apparent with the form
of this equation: the spinor field equation is real. Although the spinor
fields themselves are complex, their real and imaginary part follow a separate
equation of motion. This facilitates numerical optimisations, such as
vectorised arithmetics, and it requires a smaller memory-to-cache bandwidth.

We have to initialise the spinors in terms of eigenspinors. For this, we
diagonalise the vacuum correlation matrix
\begin{equation}
{\cal M}(\p)=\frac1{\omega_\p}\twomatrix{-\bar p_2&\bar p_1-im\\
\bar p_1+im&\bar p_2}\,.
\end{equation}
One finds that the eigenvectors are
\begin{equation}
u(\p)=\beta\twovector{Q^*\\s},\qquad
v(\p)=\beta\twovector{-s\\Q},
\end{equation}
for $p_2>0$, and
\begin{equation}
u(\p)=\beta\twovector{-s\\Q}, \qquad
v(\p)=\beta\twovector{Q^*\\s},
\end{equation}
with $Q=\bar p_1-im$, $s=\bar p_2+\omega_\p$ and $\beta^{-2}=|Q|^2+s^2$.

\section{\label{sec:doublers}Fermion doubling problem in
the semiclassical theory}

The problem of fermion doubling inevitably arises in any lattice
implementation. Since almost all numerical analyses of classical field theories
use lattice discretisation, an extension that incorporate
fermions will also share this heritage. Time discretisation, however,
is not an intrinsic parameter of the classical theory. Whereas the
space-like continuum limit simply does not exist, we can always assume that
our equations are in the time-like continuum limit. Indeed, the time-step
($a_t$) in our numerics was much smaller than the lattice spacing $a=20a_t$.

There are several remedies in the literature for the problem of doublers.
We made a version of our numerics using Wilson fermions, but the explicit
breaking of chiral symmetry introduced a linear term in
the potential. Although this can be renormalised away, not only the vacuum,
but also the physical excitations will also contribute and introduce artefacts
in the scalar effective potential. This effect will vanish in the continuum
limit, but in a semiclassical theory, we cannot go close to the continuum
limit, by construction.

The other low-cost solution could be the use of staggered fermions. These
are, however, special to two or four dimensions, and some of the doublers
will be kept. To avoid complications on the level of the equation of motion
we dropped this idea too.

In the presented numerics we simply used the naive fermion discretisation
and introduce an effective flavour number, in which we compensate for
a pair of doublers in each spatial direction. We could do this since in our
simple model there are no anomalous diagrams where doubling fermions could
cancel.

There is, however, a time-like discretisation, too, which can be a source of
time-like doublers. To eliminate
them, Aarts and Smit used a linear combination of two different flavours,
and the two degrees of freedom have both been made physical.

In the following we analyse the real-time Dirac-equation to understand
how such doublers affect our numerics.

The free Dirac propagator on spatial lattice in momentum space reads
\begin{equation}
D(t,\vec p)=\frac{m+\bar p_j\gamma^j}{2\omegap}\cos(\omega t)
-i\frac{\gamma^0}{2}\sin(\omega t)\,
\label{eq:Dslattice}
\end{equation}
with $\bar p_j=a^{-1}\sin(a p_j)$ and $\omegap^2=m^2+\sum_j {\bar p_j}^2$.
In the time-like continuum limit $\omega=\omegap$ must be satisfied so that
Eq.~(\ref{eq:Dslattice}) solves the Dirac equation. If time is discretised
as the average of the forward and backward derivative, then Dirac
equation takes the following form:
\begin{equation}
(\frac{i}{2}\gamma^0[\nabla_t^f+\nabla_t^b]+\gamma^j\bar p_j-m)D(t,\vec p)=0\,.
\label{eq:Diraclattice}
\end{equation}
Inserting Eq.~(\ref{eq:Dslattice}) into Eq.~(\ref{eq:Diraclattice})
we get following constraint: $\bar \omega=\omegap$ with
$\bar\omega=a_t^{-1}\sin(\omega a_t)$. For an extremely anisotropic
lattice ($a_t\ll a$) either $\bar\omega\approx\omega$ or
$\bar\omega\approx\pi/a_t-\omega$, since $\omegap$ is limited by the
spatial cut-off. This means, that there are two solutions (the doublers)
which can be worked out explicitly as
\begin{eqnarray}
D_1(t,\vec p)&=&\frac{m+\bar p_j\gamma^j}{2\omegap}\cos(\omegap t)
-i\frac{\gamma^0}{2}\sin(\omegap t)\,,\\
D_2(t,\vec p)&=&\frac{m+\bar p_j\gamma^j}{2\omegap}\cos(\omegap t) (-1)^s\nn
&&\quad-i\frac{\gamma^0}{2}\sin(\omegap t) (-1)^{s+1}\,,
\end{eqnarray}
where $s$ is the index of the time-slice $t$, i.e. $t=a_t s$. The sum
of these solutions is the standard lattice propagator:
\begin{equation}
D_{\rm lat}(t,\vec p)=2\left[
\frac{m+\bar p_j\gamma^j}{2\omegap}\cos(\omegap t) \chi_e(s)
-i\frac{\gamma^0}{2}\sin(\omegap t) \chi_o(s)\right]\,,
\label{eq:Dstlattice}
\end{equation}
where we introduced the $\chi_e()$ and $\chi_o()$ functions, which is one if
their integer argument is even or odd, respectively, and zero otherwise.
Indeed, Fourier transforming Eq.~(\ref{eq:Dstlattice}) yields (in the $a_t/a\ll 1$ limit)
\begin{equation}
D_{\rm lat}(p)=\pi\delta(\bar p_0^2-\omegap^2) \left[
(m+\bar p_j\gamma^j)+\gamma^0\omegap{\rm sgn}(\bar p_0)\right]\,.
\label{eq:Dplattice}
\end{equation}
We get the continuum propagator from Eq.~(\ref{eq:Dplattice}) by removing
the bars. The staggered nature of the lattice propagator is also
manifest in spatial coordinates: e.g. $\Tr D(t,\vec x)\gamma^1$ is only
then non-vanishing if $x_1/a$ is odd.

If we use the $D_{\rm lat}$ in the equations, the $\chi_e()$ function
will always give one in the source $J$, since there we close the Fermion loop
by evaluating the propagator equal space and time. At that point we need to
compensate for the extra factor two, compared to the continuum limit.
We achieve this by removing a factor of two in 
Eq.~(\ref{eq:Dstlattice}) from the initial value of $D$.

At zero time we start our system with excitations described by the $D_1$
propagator. The space and time-dependence of the background will result
an inhomogeneous propagator $D_1(x,y)$. Had we started from an initial
condition corresponding to the $D_2$ propagator, the
evolution would have lead to $D_2(x,y)$. Inserting
$D_{\rm lat}=D_1(x,y)+D_2(x,y)$
or $\bar{D}_{\rm lat}=D_1(x,y)-D_2(x,y)$
into the inhomogeneous Dirac equation one discovers
that these linear combinations decouple: they communicate only through
the back-reaction to the scalars. The various Lorentz-components of
$D_{\rm lat}(x,y)$ and $\bar{D}_{\rm lat}(x,y)$ couple to the background at
different time slices, depending on the parity of $x^0-y^0$.
If the background is a smooth function of time, $D_{\rm lat}$ and
$\bar{D}_{\rm lat}$ will evolve on the same background, up to an error
$\sim a_t$.
Thus, their difference, $D_2(x,y)$ is suppressed by the time-like spacing,
i.e. if there is no $D_2$ component in our initial condition, the production
of doubler particles will be small compared to the standard particles. In
the back-reaction and the measured spectra both types of excitations
contribute indistinguishably. (Notice that $D_1(t,\vec p)$ and $D_2(t,\vec p)$
are identical at equal time, where these observables are taken.)
Similar ideas have been implemented to tackle the species doubling problem in
the context of the hard thermal loop effective action of the electroweak theory
in Ref.~\cite{DietrichGuyKari}.

To check these ideas we plotted the damping of the scalar field in
Fig.~\ref{fig:damping}. For this calculation in 2+1 dimensions we used
the effectve flavours number $1/4$. An erronous estimation of the
number of flavours should have generated an unexpected factor 2 in the
rate.


\end{document}